\documentclass[aps,prl,twocolumn,groupedaddress,showpacs]{revtex4}

\usepackage{graphicx}

\begin{document}

\title{Azimuthally polarized spatial dark solitons: exact solutions of Maxwell's equations in a Kerr medium}

\author{Alessandro Ciattoni}
\email{alessandro.ciattoni@aquila.infn.it} \affiliation{Istituto Nazionale per la Fisica della Materia, UdR
Universit\'a dell'Aquila, 67010, L'Aquila, Italy, and
\\ Dipartimento di Fisica, Universit\'a dell'Aquila, 67010 L'Aquila, Italy}

\author{Bruno Crosignani}
\affiliation{Dipartimento di Fisica, Universit\'a dell'Aquila, 67010 L'Aquila, Italy and \\ Istituto Nazionale di
Fisica della Materia, UdR Roma "La Sapienza", 00185 Roma, Italy and \\ California Institute of Technology 128-95,
Pasadena, California 91125}

\author{Paolo Di Porto}
\affiliation{Dipartimento di Fisica, Universit\'a dell'Aquila, 67010, L'Aquila, Italy and \\ Istituto Nazionale di
Fisica della Materia, UdR Roma "La Sapienza", 00185 Roma, Italy}

\author{Amnon Yariv}
\address{California Institute of Technology 128-95, Pasadena, California 91125}

\date{\today}

\begin{abstract}
Spatial Kerr solitons, typically associated with the standard paraxial nonlinear Schr\"{o}dinger equation, are
shown to exist to all nonparaxial orders, as exact solutions of Maxwell's equations in the presence of vectorial
Kerr effect. More precisely, we prove the existence of {\it azimuthally polarized}, spatial, dark soliton
solutions of Maxwell's equations, while exact linearly polarized (2+1)-D solitons do not exist. Our {\it ab
initio} approach predicts the existence of dark solitons up to an upper value of the maximum field amplitude,
corresponding to a minimum soliton width of about one fourth of the wavelength.
\end{abstract}


\maketitle

Bright and dark optical spatial solitons have been and still are the object of an intense theoretical and
experimental investigation. \cite{B1,B2} In particular, Kerr solitons have always represented the theoretical
reference model in view of the simplicity of their analytic description. This is mainly due to the fact that they
are solution of an equation, the so-called nonlinear Schr\"{o}dinger equation (NLS), which is exactly integrable
and also admits, in particular cases, of simple analytic solutions. The main limit of the NLS is that it is only
able to describe scalar optical propagation in the paraxial approximation. Nonparaxial descriptions have been
introduced by adopting suitable asymptotic expansions in the smallness parameter $\lambda / d$ (where $d$ is the
beam width) which result in improved versions of NLS including both scalar \cite{sca1,sca2,sca3,sca4} and
vectorial \cite{vec1,vec2,vec3} contributions. Besides eliminating the unphysical catastrophic collapse associated
with standard NLS \cite{coq1,coq2,coq3}, they show that the transverse Cartesian components of (2+1)-D propagating
beams are mutually coupled, so that linearly polarized (2+1)-D Kerr solitons cannot exist. The same nonparaxial
approach can also be used to predict the existence of (1+1)-D bright and dark solitons \cite{NPB1,NPB2}, to the
first significant order in $\lambda / d$.

The standard or improved versions of the NLS are limited by the underlying approximation scheme and fail whenever
$\lambda / d$ approaches one. The question naturally arises: are there solitons to all nonparaxial orders? More in
general, is it possible to find exact soliton solutions of Maxwell equations?

In this paper we find, by starting from Maxwell's equations, that an exact solution (to all nonparaxial orders)
indeed exists in the form of an azimuthally polarized dark soliton, i.e.
\begin{equation} \label{sol1}
{\bf E}(r,\varphi,z) = e^{i \chi z } E(r)  \hat{\bf e}_\varphi,
\end{equation}
where $r,\varphi,z$ are cylindrical coordinates with unit vectors $ \hat{\bf e}_r,  \hat{\bf e}_\varphi,  \hat{\bf
e}_z$ and $\chi$ is a real number. More precisely, we are able to prove the existence of azimuthally polarized
dark solitons, i.e., of fields of the kind given in Eq.(\ref{sol1}), where $E(r)$ monothonically ranges from
$E(0)=0$ to $E(\infty) = E_\infty$ and $E_\infty$ is a given real constant. We numerically evaluate the soliton
shape and its existence curve, our results being obtained {\it without any approximation}, so that the whole range
of possible soliton widths is considered without any formal distinction between paraxial and nonparaxial regimes.
We note that the existence of azimuthally polarized, circularly symmetric fields of the form $E(r,z) \hat{\bf
e}_\varphi$, of which the soliton given in Eq.(\ref{sol1}) is a specific case, is due to the simple symmetry
properties of Kerr nonlinearity.

A monochromatic electromagnetic field $Re[{\bf E} \exp(-i \omega t)]$, $Re[{\bf B} \exp(-i \omega t)]$ propagating
in a nonlinear medium obeys Maxwell's equations
\begin{eqnarray} \label{maxwell1}
\nabla \times {\bf E} &=& i \omega {\bf B}, \nonumber \\
\nabla \times {\bf B} &=& -i \frac{\omega}{c^2} n_0^2 {\bf E} - i \omega \mu_0 {\bf P}_{nl},
\end{eqnarray}
where $n_0$ labels the linear refractive index and ${\bf P}_{nl}$ is the nonlinear polarizability. In the case of
nonresonant isotropic media \cite{Cros}, the vectorial Kerr effect is described by the polarizability
\begin{equation} \label{polariz}
{\bf P}_{nl} = \frac{4}{3} \epsilon_0 n_0 n_2 \left[ |{\bf E}|^2 {\bf E} + \frac{1}{2} ({\bf E \cdot E}) {\bf E}^*
\right],
\end{equation}
$n_2$ being the nonlinear refractive index coefficient. After eliminating $\bf B$ from Eq.(\ref{maxwell1}) we get
\begin{equation} \label{maxwell2}
\nabla \times \nabla \times {\bf E} =  k^2 {\bf E} + k^2 \frac{4}{3} \frac{n_2}{n_0}  \left[ |{\bf E}|^2 {\bf E} +
\frac{1}{2} ({\bf E \cdot E}) {\bf E}^* \right]
\end{equation}
where $k = n_0 \omega/c$. Let us consider fields of the form
\begin{equation} \label{field}
{\bf E}(r,\varphi\,z) = E_\varphi (r,z) \hat{\bf e}_\varphi + E_z(r,z) \hat{\bf e}_z,
\end{equation}
describing a circularly symmetric field with vanishing radial component. Inserting Eq.(\ref{field}) in
Eq.(\ref{maxwell2}), we obtain
\begin{eqnarray} \label{set}
&& \frac{\partial^2 E_z}{\partial r \partial z} = 0, \nonumber \\
&& \frac{\partial ^2 E_\varphi}{\partial z^2} + \frac{\partial}{\partial r} \left( \frac{\partial
E_\varphi}{\partial r} +\frac{E_\varphi}{r} \right) = -k^2 E_\varphi \nonumber \\
&& - k^2 \frac{4}{3} \frac{n_2}{n_0}  \left[ |{\bf E}|^2 E_\varphi + \frac{1}{2}
({\bf E \cdot E}) E_\varphi^* \right], \nonumber \\
&& \frac{\partial ^2 E_z}{\partial r^2} + \frac{1}{r} \frac{\partial E_z}{\partial r} = -k^2 E_z \nonumber \\
&& - k^2 \frac{4}{3} \frac{n_2}{n_0}  \left[ |{\bf E}|^2 E_z + \frac{1}{2} ({\bf E \cdot E}) E_z^* \right].
\end{eqnarray}
Internal consistency of the set of Eqs.(\ref{set}) (three equations in two unknowns) requires $E_z = 0$. As a
consequence, the second of Eqs.(\ref{set}) yields
\begin{equation} \label{prop}
\frac{\partial ^2 E_\varphi}{\partial z^2} + \frac{\partial}{\partial r} \left( \frac{\partial E_\varphi}{\partial
r} +\frac{E_\varphi}{r} \right) =  -k^2 E_\varphi -2 k^2 \frac{n_2}{n_0}  |E_\varphi|^2 E_\varphi.
\end{equation}
We note that circular symmetry and polarization imposed to the field, together with the symmetry properties of
Kerr effect, have allowed us to reduce Maxwell's equations to the single Eq.(\ref{prop}). Equation (\ref{prop}) is
conveniently rewritten in the dimensionless form
\begin{equation} \label{adim}
\frac{\partial ^2 U}{\partial \zeta^2} + 2 \frac{\partial}{\partial \rho} \left( \frac{\partial U}{\partial \rho}
+\frac{U}{\rho} \right) =  -U - 2 \gamma |U|^2 U,
\end{equation}
where $\rho=\sqrt{2} k r$, $\zeta = kz$, $U =\sqrt{|n_2|/n_0} E_\varphi$ and $\gamma = n_2 / |n_2|$. If we look
for soliton solutions of the form
\begin{equation} \label{sol2}
U(\rho,\zeta) = e^{i \alpha \zeta} u(\rho),
\end{equation}
where (see Eq.(\ref{sol1})) $\alpha = \chi /k$, Eq.(\ref{adim}) becomes
\begin{equation} \label{eqsol}
\frac{d}{d \rho}\left( \frac{d u}{d \rho} + \frac{u}{\rho} \right) = \frac{1}{2}(\alpha^2-1) u - \gamma u ^3.
\end{equation}
Both the structure of Eq.(\ref{eqsol}) and the azimuthal field polarization dictate $u(0)=0$, so that azimuthally
polarized bright solitons do not exist. In order to find dark solitons, we introduce the further condition
\begin{equation}
\lim_{\rho \rightarrow \infty} u(\rho) = u_\infty,
\end{equation}
together with the vanishing of all derivatives for $\rho \rightarrow \infty$. Since focusing media ($\gamma = 1$,
i.e., $n_2>0$) are not able to support dark solitons, we consider hereafter defocusing media ($\gamma = -1$, i.e.,
$n_2<0$), so that Eq.(\ref{eqsol}) reads
\begin{equation} \label{eqdef}
\frac{d}{d \rho}\left( \frac{d u}{d \rho} + \frac{u}{\rho} \right) = \frac{1}{2}(\alpha^2-1) u + u ^3,
\end{equation}
which implies, together with the above boundary condition at infinity,
\begin{equation} \label{alf}
\alpha = \pm \sqrt{1-2u_\infty^2}.
\end{equation}
\begin{figure}
\includegraphics[width=0.4\textwidth]{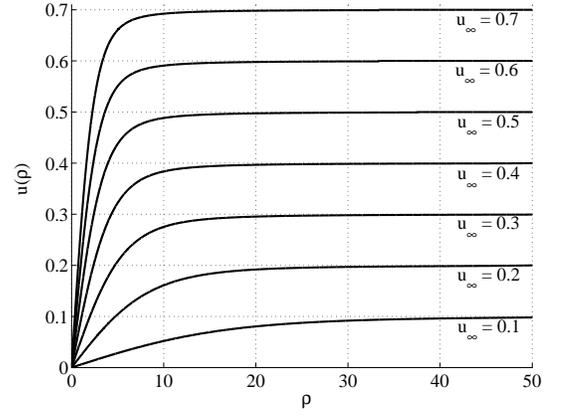}
\caption{Soliton profile $u(\rho)$ for various values of $u_\infty$.} \label{F1}
\end{figure}
\begin{figure}
\includegraphics[width=0.4\textwidth]{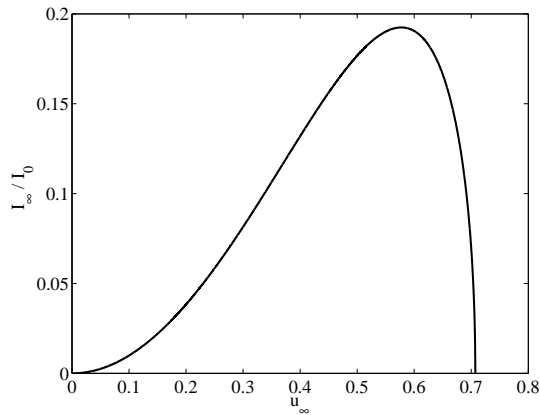}
\caption{Normalized asymptotic optical intensity $I_\infty / I_0$ as a function of the asymptotic adimensional
field amplitude $u_\infty$. Note that two solitons exist for any allowed asymptotic optical intensity.} \label{F2}
\end{figure}
While positive and negative signs of $\alpha$ respectively refer to forward and backward travelling solitons (see
Eq.(\ref{sol2})), $u(\rho)$ depends on $\alpha^2$ (see Eq.(\ref{eqsol})) and is clearly the same in the two cases.
Equation (\ref{alf}) shows the existence of an upper threshold for the soliton asymptotic amplitude
\begin{equation} \label{threshold}
u_\infty < \frac{1}{\sqrt{2}},
\end{equation}
since, otherwise, $\alpha$ would become imaginary. From an intuitive point of view, the existence of this
threshold is related to the dominant defocusing effect due to the nonlinearity over the focusing one due to
diffraction. If we now insert Eq.(\ref{alf}) into Eq.(\ref{eqdef}), we obtain
\begin{equation} \label{eqsol1}
\frac{d}{d \rho}\left( \frac{d u}{d \rho} + \frac{u}{\rho} \right) = (u^2-u_\infty^2) u.
\end{equation}

We have carried out a numerical integration of Eq.(\ref{eqsol1}) with boundary conditions $u(0)=0$ and
$u(\infty)=u_\infty$, by employing a standard shooting-relaxation method for boundary value problems. The results
of our simulations show that dark solitons can be obtained in the range of field amplitudes $0 < u_\infty <
1/\sqrt{2}$. Different soliton profiles are reported in Fig.1.

In order to complete our analysis, we now evaluate both the magnetic field and the Poynting vector. Recalling the
expression of the soliton electric field
\begin{equation} \label{ElecField}
{\bf E} = \sqrt{\frac{n_0}{|n_2|}} e^{i \alpha k z } u(\sqrt{2}kr)  \hat{\bf e}_\varphi,
\end{equation}
we obtain, from the first of Eqs.(\ref{maxwell1}) written in cylindrical coordinates,
\begin{equation} \label{MagField}
{\bf B} = - \sqrt{\frac{n_0}{|n_2|}} e^{i \alpha k z } \frac{k}{\omega}
         \left[ \alpha u \hat{\bf e}_r + i \sqrt{2} \left( \frac{du}{d\rho} + \frac{u}{\rho} \right)
         \hat{\bf e}_z \right]_{\rho=\sqrt{2}kr}.
\end{equation}
The magnetic field has a radial component whose shape coincides with that of the electric field, and a vanishing
azimuthal component, so that $\bf E$ and $\bf B$ are mutually orthogonal. With the help of Eqs.(\ref{ElecField})
and (\ref{MagField}), the time averaged Poynting vector
\begin{equation} \label{Poy}
{\bf S} = \frac{1}{2 \mu_0} Re \left( {\bf E} \times {\bf B}^* \right)
\end{equation}
turns out to be given by
\begin{equation} \label{Poyn}
{\bf S} (r) =  \frac{\alpha k}{2 \omega \mu_0} \frac{n_0}{|n_2|} u^2(\sqrt{2}kr) \hat{\bf e}_z = \frac{\alpha k}{2
\omega \mu_0} |{\bf E}|^2 \hat{\bf e}_z.
\end{equation}
\begin{figure}
\includegraphics[width=0.4\textwidth]{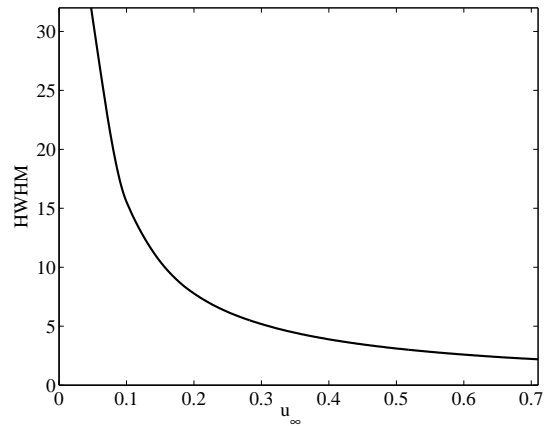}
\caption{Existence curve relating the normalized soliton optical intensity HWHM to $u_\infty$.} \label{F3}
\end{figure}
We note that $\bf S$ is parallel to the $z-$ axis, consistently with the shape-invariant nature of solitons. From
an analytical point of view, this corresponds to the $\pi /2$ phase difference between $B_z$ and $E_\varphi$ (see
Eqs.(\ref{ElecField}) and (\ref{MagField})). As expected, the Poynting vector is either parallel or antiparallel
to $\hat{\bf e}_z$ according to the sign of $\alpha$, while its amplitude is proportional to $|{\bf E}|^2$. The
above plane wave-like properties are consistent with the nondiffractive nature of exact solitons.

It is worthwhile to underline that, in the case of the azimuthal dark solitons we are considering, the asymptotic
optical intensity $I_\infty = |{\bf S(\infty)}|$ turns out not to be proportional to $u_\infty^2$. In fact, by
using Eqs.(\ref{alf}) and (\ref{Poyn}), one obtains
\begin{equation} \label{Inw}
I_\infty(u_\infty) = I_0 u_\infty^2 \sqrt{1-2u_\infty^2}
\end{equation}
where $I_0 = kn_0 /(2\omega \mu_0 |n_2|)$, whose profile is reported in Fig.2. Equation (\ref{Inw}) shows that the
asymptotic optical intensity is not a monotonically increasing function of the asymptotic field amplitude, but
reaches its maximum threshold value $I_\infty^{max} = I_0 / 3^{3/2}$ in correspondence to $u_\infty = 1/\sqrt{3}$.
This is connected to the $\alpha-$ dependence of the magnetic field (see Eq.(\ref{MagField})) whose radial part
tends to vanish for $u_\infty \rightarrow 1/\sqrt{2}$. A related and relevant consequence of Eq.(\ref{Inw}) is the
existence of {\it two solitons} of different widths for a given asymptotic optical intensity. The existence curve
relating the normalized half width at half maximum (HWHM) of the soliton optical intensity profile $|{\bf
S}(\rho)|$ to $u_\infty$ is reported in Fig.3. In particular, Fig.3 shows the existence of a normalized minimum
HWHM $\simeq 2.1$ ($\simeq 0.24 \lambda$) for $u_\infty = 1/\sqrt{2}$. In addition, Fig.2 shows that a normalized
HWHM $\simeq 2.7$ ($\simeq 0.3 \lambda$) corresponds to $u_\infty = 1/\sqrt{3}$ for which the soliton attains the
maximum asymptotic optical intensity $I_\infty^{max}$.

\begin{figure}
\includegraphics[width=0.4\textwidth]{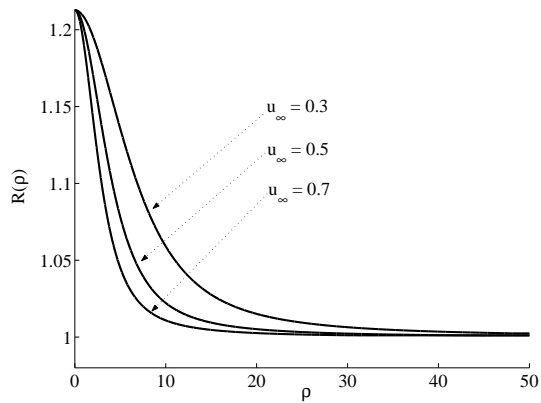}
\caption{Plot of the ratio $R(\rho) = u_\infty \tanh (u_\infty \rho/\sqrt{2}) / u(\rho)$ for different values of
$u_\infty$. } \label{F1}
\end{figure}

It is interesting to examine the behavior of our solution in the limit of large $\rho$. To this end, neglecting in
Eq.(\ref{eqsol1}) the term in $u/\rho$, we have
\begin{equation} \label{nls}
\frac{d^2 u}{d \rho^2} = (u^2 - u_\infty^2) u
\end{equation}
which formally coincides with the equation describing one dimensional linearly polarized paraxial dark solitons.
Equation (\ref{nls}) admits of the solution $u= u_\infty \tanh (\rho u_\infty /\sqrt{2})$. This solution can be
compared with the exact one. This is done in Fig.4 where the ratio $R(\rho)$ between the hyperbolic tangent and
the exact solution is reported as function of $\rho$, for different values of $u_\infty$. The hyperbolic tangent
solution reproduces the exact one for large values of $\rho$, as expected, while it at most differs by a factor
$\cong 1.2$ for small values of $\rho$.

Recent findings indicate that radially and azimuthally polarized laser beams can be efficiently generated
\cite{azi1,azi2} and focused to subwavelength diameter \cite{azi3} in resonator-like schemes. These, combined with
an appropriate nonlinear material can form a basis for a first experimental investigation of the existence of the
azimuthally polarized solitons. As an example, if we consider a strongly defocusing nonlinear medium, like Sodium
vapor at a density of $\sim 10^{12} atoms/cm^3$ for which $n_2 = -4 \cdot 10^{-14} (m/Volt)^2$ \cite{sodium}, the
upper threshold for the asymptotic optical soliton intensity $I_\infty^{max}$ is of the order of $.5 \: MW /
cm^2$.

In summary, we have found, to the best of our knowledge, the first exact, two-dimensional, dark spatial
Kerr-soliton. This soliton is obtained as a particular solution of an equation (Eq.(\ref{adim})), which in turn
exactly follows from Maxwell's equations, describing circularly symmetric, azimuthally polarized electric fields.
Our {\it ab initio} approach inherently overcomes any approximated scheme employed to describe paraxial or
slightly nonparaxial situations. In particular, the derived analytical expression of the soliton propagation
constant as a function of the normalized ($\rho \rightarrow \infty$) field amplitude $u_\infty$ shows that the
soliton does not exist whenever $u_\infty > 1/\sqrt{2}$. We have also numerically evaluated the existence curve
relating the soliton width to $u_\infty$; as a relevant result, this curve shows that the soliton width attains
the minimum possible value of about $\lambda/4$ for $u_\infty$ approaching $1/\sqrt{2}$.

This research has been funded by the Istituto Nazionale di Fisica della Materia through the "Solitons embedded in
holograms", the FIRB "Space-Time nonlinear effects" projects and the Air Force Office of Scientific Research (H.
Schlossberg).

\bibliography{references}

\begin{thebibliography}{19}
\expandafter\ifx\csname natexlab\endcsname\relax\def\natexlab#1{#1}\fi
\expandafter\ifx\csname bibnamefont\endcsname\relax
  \def\bibnamefont#1{#1}\fi
\expandafter\ifx\csname bibfnamefont\endcsname\relax
  \def\bibfnamefont#1{#1}\fi
\expandafter\ifx\csname citenamefont\endcsname\relax
  \def\citenamefont#1{#1}\fi
\expandafter\ifx\csname url\endcsname\relax
  \def\url#1{\texttt{#1}}\fi
\expandafter\ifx\csname urlprefix\endcsname\relax\def\urlprefix{URL }\fi
\providecommand{\bibinfo}[2]{#2}
\providecommand{\eprint}[2][]{\url{#2}}

\bibitem[{\citenamefont{Trillo and Torruellas}(2001)}]{B1}
\bibinfo{author}{\bibfnamefont{S.}~\bibnamefont{Trillo}} \bibnamefont{and}
  \bibinfo{author}{\bibfnamefont{W.}~\bibnamefont{Torruellas}},
  \emph{\bibinfo{title}{Spatial Solitons}} (\bibinfo{publisher}{Springer},
  \bibinfo{address}{Berlin}, \bibinfo{year}{2001}).

\bibitem[{\citenamefont{Kivshar and Luther-Davis}(1998)}]{B2}
\bibinfo{author}{\bibfnamefont{Y.}~\bibnamefont{Kivshar}} \bibnamefont{and}
  \bibinfo{author}{\bibfnamefont{B.}~\bibnamefont{Luther-Davis}},
  \bibinfo{journal}{Phys. Reports} \textbf{\bibinfo{volume}{298}},
  \bibinfo{pages}{81} (\bibinfo{year}{1998}).

\bibitem[{\citenamefont{Feit and J.~A.~Fleck}(1998)}]{sca1}
\bibinfo{author}{\bibfnamefont{M.~D.} \bibnamefont{Feit}} \bibnamefont{and}
  \bibinfo{author}{\bibfnamefont{J.}~\bibnamefont{J.~A.~Fleck}},
  \bibinfo{journal}{J. Opt. Soc. Am. B} \textbf{\bibinfo{volume}{5}},
  \bibinfo{pages}{633} (\bibinfo{year}{1998}).

\bibitem[{\citenamefont{Akhmediev et~al.}(1993)\citenamefont{Akhmediev,
  Ankiewicz, and Soto-Crespo}}]{sca2}
\bibinfo{author}{\bibfnamefont{N.}~\bibnamefont{Akhmediev}},
  \bibinfo{author}{\bibfnamefont{A.}~\bibnamefont{Ankiewicz}},
  \bibnamefont{and} \bibinfo{author}{\bibfnamefont{J.~M.}
  \bibnamefont{Soto-Crespo}}, \bibinfo{journal}{Opt. Lett.}
  \textbf{\bibinfo{volume}{15}}, \bibinfo{pages}{411} (\bibinfo{year}{1993}).

\bibitem[{\citenamefont{Fibich}(1996)}]{sca3}
\bibinfo{author}{\bibfnamefont{G.}~\bibnamefont{Fibich}},
  \bibinfo{journal}{Phys. Rev. Lett.} \textbf{\bibinfo{volume}{76}},
  \bibinfo{pages}{4356} (\bibinfo{year}{1996}).

\bibitem[{\citenamefont{Sheppard and Haelterman}(1998)}]{sca4}
\bibinfo{author}{\bibfnamefont{A.~P.} \bibnamefont{Sheppard}} \bibnamefont{and}
  \bibinfo{author}{\bibfnamefont{M.}~\bibnamefont{Haelterman}},
  \bibinfo{journal}{Opt. Lett.} \textbf{\bibinfo{volume}{23}},
  \bibinfo{pages}{1820} (\bibinfo{year}{1998}).

\bibitem[{\citenamefont{Chi and Guo}(1995)}]{vec1}
\bibinfo{author}{\bibfnamefont{S.}~\bibnamefont{Chi}} \bibnamefont{and}
  \bibinfo{author}{\bibfnamefont{Q.}~\bibnamefont{Guo}}, \bibinfo{journal}{Opt.
  Lett.} \textbf{\bibinfo{volume}{20}}, \bibinfo{pages}{1598}
  (\bibinfo{year}{1995}).

\bibitem[{\citenamefont{Fibich and Ilan}(2001)}]{vec2}
\bibinfo{author}{\bibfnamefont{G.}~\bibnamefont{Fibich}} \bibnamefont{and}
  \bibinfo{author}{\bibfnamefont{B.}~\bibnamefont{Ilan}},
  \bibinfo{journal}{Physica D} \textbf{\bibinfo{volume}{157}},
  \bibinfo{pages}{112} (\bibinfo{year}{2001}).

\bibitem[{\citenamefont{Ciattoni et~al.}(2002)\citenamefont{Ciattoni, Conti,
  DelRe, Porto, Crosignani, and Yariv}}]{vec3}
\bibinfo{author}{\bibfnamefont{A.}~\bibnamefont{Ciattoni}},
  \bibinfo{author}{\bibfnamefont{C.}~\bibnamefont{Conti}},
  \bibinfo{author}{\bibfnamefont{E.}~\bibnamefont{DelRe}},
  \bibinfo{author}{\bibfnamefont{P.~D.} \bibnamefont{Porto}},
  \bibinfo{author}{\bibfnamefont{B.}~\bibnamefont{Crosignani}},
  \bibnamefont{and} \bibinfo{author}{\bibfnamefont{A.}~\bibnamefont{Yariv}},
  \bibinfo{journal}{Opt. Lett.} \textbf{\bibinfo{volume}{27}},
  \bibinfo{pages}{734} (\bibinfo{year}{2002}).

\bibitem[{\citenamefont{Silberberg}(1990)}]{coq2}
\bibinfo{author}{\bibfnamefont{Y.}~\bibnamefont{Silberberg}},
  \bibinfo{journal}{Opt. Lett.} \textbf{\bibinfo{volume}{15}},
  \bibinfo{pages}{1282} (\bibinfo{year}{1990}).

\bibitem[{\citenamefont{Fibich and Gaeta}(2000)}]{coq3}
\bibinfo{author}{\bibfnamefont{G.}~\bibnamefont{Fibich}} \bibnamefont{and}
  \bibinfo{author}{\bibfnamefont{A.~L.} \bibnamefont{Gaeta}},
  \bibinfo{journal}{Opt. Lett.} \textbf{\bibinfo{volume}{25}},
  \bibinfo{pages}{335} (\bibinfo{year}{2000}).

\bibitem[{\citenamefont{Kelly}(1965)}]{coq1}
\bibinfo{author}{\bibfnamefont{P.}~\bibnamefont{Kelly}},
  \bibinfo{journal}{Phys. Rev. Lett.} \textbf{\bibinfo{volume}{15}},
  \bibinfo{pages}{1500} (\bibinfo{year}{1965}).

\bibitem[{\citenamefont{Crosignani et~al.}(2004)\citenamefont{Crosignani,
  Yariv, and Mookherjea}}]{NPB1}
\bibinfo{author}{\bibfnamefont{B.}~\bibnamefont{Crosignani}},
  \bibinfo{author}{\bibfnamefont{A.}~\bibnamefont{Yariv}}, \bibnamefont{and}
  \bibinfo{author}{\bibfnamefont{S.}~\bibnamefont{Mookherjea}},
  \bibinfo{journal}{Opt. Lett.} \textbf{\bibinfo{volume}{29}},
  \bibinfo{pages}{1254} (\bibinfo{year}{2004}).

\bibitem[{\citenamefont{Ciattoni et~al.}()\citenamefont{Ciattoni, Crosignani,
  Yariv, and Mookherjea}}]{NPB2}
\bibinfo{author}{\bibfnamefont{A.}~\bibnamefont{Ciattoni}},
  \bibinfo{author}{\bibfnamefont{B.}~\bibnamefont{Crosignani}},
  \bibinfo{author}{\bibfnamefont{A.}~\bibnamefont{Yariv}}, \bibnamefont{and}
  \bibinfo{author}{\bibfnamefont{S.}~\bibnamefont{Mookherjea}},
  \bibinfo{journal}{Opt. Lett. (to be published)}  (????).

\bibitem[{\citenamefont{Crosignani et~al.}(1982)\citenamefont{Crosignani,
  Cutolo, and Porto}}]{Cros}
\bibinfo{author}{\bibfnamefont{B.}~\bibnamefont{Crosignani}},
  \bibinfo{author}{\bibfnamefont{A.}~\bibnamefont{Cutolo}}, \bibnamefont{and}
  \bibinfo{author}{\bibfnamefont{P.~D.} \bibnamefont{Porto}},
  \bibinfo{journal}{J. Opt. Soc. Am.} \textbf{\bibinfo{volume}{72}},
  \bibinfo{pages}{1136} (\bibinfo{year}{1982}).

\bibitem[{\citenamefont{Oron et~al.}(2000)\citenamefont{Oron, Bilt, Davidson,
  Friesem, Bomzon, and Hasman}}]{azi1}
\bibinfo{author}{\bibfnamefont{R.}~\bibnamefont{Oron}},
  \bibinfo{author}{\bibfnamefont{S.}~\bibnamefont{Bilt}},
  \bibinfo{author}{\bibfnamefont{N.}~\bibnamefont{Davidson}},
  \bibinfo{author}{\bibfnamefont{A.~A.} \bibnamefont{Friesem}},
  \bibinfo{author}{\bibfnamefont{Z.}~\bibnamefont{Bomzon}}, \bibnamefont{and}
  \bibinfo{author}{\bibfnamefont{E.}~\bibnamefont{Hasman}},
  \bibinfo{journal}{Phys. Rev. Lett.} \textbf{\bibinfo{volume}{77}},
  \bibinfo{pages}{3322} (\bibinfo{year}{2000}).

\bibitem[{\citenamefont{Moshe et~al.}(2003)\citenamefont{Moshe, Jackel, and
  Meir}}]{azi2}
\bibinfo{author}{\bibfnamefont{I.}~\bibnamefont{Moshe}},
  \bibinfo{author}{\bibfnamefont{S.}~\bibnamefont{Jackel}}, \bibnamefont{and}
  \bibinfo{author}{\bibfnamefont{A.}~\bibnamefont{Meir}},
  \bibinfo{journal}{Opt. Lett.} \textbf{\bibinfo{volume}{28}},
  \bibinfo{pages}{807} (\bibinfo{year}{2003}).

\bibitem[{\citenamefont{Dorn et~al.}(2003)\citenamefont{Dorn, Quabis, and
  Leuchs}}]{azi3}
\bibinfo{author}{\bibfnamefont{R.}~\bibnamefont{Dorn}},
  \bibinfo{author}{\bibfnamefont{S.}~\bibnamefont{Quabis}}, \bibnamefont{and}
  \bibinfo{author}{\bibfnamefont{G.}~\bibnamefont{Leuchs}},
  \bibinfo{journal}{Phys. Rev. Lett.} \textbf{\bibinfo{volume}{91}},
  \bibinfo{pages}{233901} (\bibinfo{year}{2003}).

\bibitem[{\citenamefont{Swartzlander et~al.}(1991)\citenamefont{Swartzlander,
  Andersen, Regan, Yin, and Kaplan}}]{sodium}
\bibinfo{author}{\bibfnamefont{G.~A.} \bibnamefont{Swartzlander}},
  \bibinfo{author}{\bibfnamefont{D.~R.} \bibnamefont{Andersen}},
  \bibinfo{author}{\bibfnamefont{J.~J.} \bibnamefont{Regan}},
  \bibinfo{author}{\bibfnamefont{H.}~\bibnamefont{Yin}}, \bibnamefont{and}
  \bibinfo{author}{\bibfnamefont{A.}~\bibnamefont{Kaplan}},
  \bibinfo{journal}{Phys. Rev. Lett.} \textbf{\bibinfo{volume}{66}},
  \bibinfo{pages}{1583} (\bibinfo{year}{1991}).

\end{thebibliography}
\end{document}